\documentclass[twocolumn,showpacs,superscriptaddress,showkeys,reprint,aps]{revtex4-1}

\usepackage{graphicx}
\usepackage{dcolumn}
\usepackage{bm}
\usepackage{tabularx}
\usepackage{amsmath}
\usepackage{amssymb}
\usepackage{enumitem}
\usepackage{multirow}
\usepackage[bookmarks,bookmarksnumbered]{hyperref}
\hypersetup{colorlinks = true,
            linkcolor = blue,
            anchorcolor =red,
            citecolor = blue,
            pdfauthor=author}
\usepackage[capitalise]{cleveref}
\usepackage{xcolor}

\begin{document}

\title{Fourier-Flow model generating Feynman paths}
\date{\today}

\author{Shile Chen}
    \affiliation{Department of Physics, Tsinghua University, Beijing 100084, China.}

\author{Oleh Savchuk}
    \affiliation{Frankfurt Institute for Advanced Studies, Ruth Moufang Strasse 1, D-60438, Frankfurt am Main, Germany}

   \affiliation{GSI Helmholtzzentrum f\"ur Schwerionenforschung GmbH, D-64291 Darmstadt, Germany}

    \affiliation{Bogolyubov Institute for Theoretical Physics, 03680 Kyiv, Ukraine}
    
\author{Shiqi Zheng}
    \affiliation{University of North Carolina at Chapel Hill, North Carolina 27599, USA}

   \affiliation{Department of Physics, Tianjin University, Tianjin 300354, China}
   
\author{Baoyi Chen}
   \affiliation{Department of Physics, Tianjin University, Tianjin 300354, China}
   
\author{Horst Stoecker}
    \affiliation{Frankfurt Institute for Advanced Studies, Ruth Moufang Strasse 1, D-60438, Frankfurt am Main, Germany}
    \affiliation{GSI Helmholtzzentrum f\"ur Schwerionenforschung GmbH, D-64291 Darmstadt, Germany}
    \affiliation{Institut f\"ur Theoretische Physik, Goethe Universit\"at Frankfurt, D-60438 Frankfurt am Main, Germany}

\author{Lingxiao Wang}
\email{lwang@fias.uni-frankfurt.de}
    \affiliation{Frankfurt Institute for Advanced Studies, Ruth Moufang Strasse 1, D-60438, Frankfurt am Main, Germany}

\author{Kai Zhou}
\email{zhou@fias.uni-frankfurt.de}
    \affiliation{Frankfurt Institute for Advanced Studies, Ruth Moufang Strasse 1, D-60438, Frankfurt am Main, Germany}

\begin{abstract}
As an alternative but unified and more fundamental description for quantum physics, Feynman path integrals generalize the classical action principle to a probabilistic perspective, under which the physical observables' estimation translates into a weighted sum over all possible paths. The underlying difficulty is to tackle the whole path manifold from finite samples that can effectively represent the Feynman propagator dictated probability distribution. Modern generative models in machine learning can handle learning and representing probability distribution with high computational efficiency. In this study, we propose a Fourier-flow generative model to simulate the Feynman propagator and generate paths for quantum systems. As demonstration, we validate the path generator on the harmonic and anharmonic oscillators. The latter is a double-well system without analytic solutions. To preserve the periodic condition for the system, the Fourier transformation is introduced into the flow model to approach a Matsubara representation. With this novel development, the ground-state wave function and low-lying energy levels are estimated accurately. Our method offers a new avenue to investigate quantum systems with machine learning assisted Feynman Path integral solving.
\end{abstract}

\maketitle
\section{Introduction}
Feynman path integrals\cite{Feynman:1948ur,feynman2010quantum} have been proven successful in describing quantum systems, which include contributions from both the classical path and quantum fluctuations. For physical observables' estimation of a quantum system, the functional path integral turns it to be an ergodic problem, since it requires the summation over an infinite number of quantum-mechanically possible trajectories. In such a perspective, except for few analytically solvable systems, it's routinely needed to generate paths $x(t)$ following distribution $e^{iS[x(t)]/\hbar}$ given an action $S[x(t)]$. To sample probable paths, the Markov Chain Monte Carlo (MCMC) technique is widely adopted as a traditional numerical approach~\cite{Carlson:2014vla,Westbroek:2017tym}. However, time-consuming updates inevitably emerge in MCMC when one attempts to propose uncorrelated paths for 
a large system. It thus calls for effective novel path generation methods, where machine learning generative algorithms could be introduced~\cite{Boyda:2022nmh}.

Generative models of machine learning~\cite{Kingma:2013hel,goodfellow2014generative} have been shown particularly useful in capturing the underlying probability distributions hidden in data or explicitly existing in physical systems~\cite{Zhou:2018ill,Pawlowski:2018qxs,Wang:2020hji}. Normalizing flows\cite{rezende2015variational}, starting from a plain prior distribution, is able to evolve to desired distribution through bijective transformations. These transformations can be constructed via neural networks, making them flexible enough due to the universal approximation theorem. Combined with MCMC, normalizing flows offer a traceable and efficient way for sampling from a target distribution, which is currently burgeoning in lattice QFT studies~\cite{2020arXiv201201442M,Albergo:2019eim,Nicoli:2020njz,Boyda:2022nmh,Caselle:2022acb}.

To advance the effectiveness of generative models for physics, introducing corresponding constraints or special network architectures would be crucial, e.g., embedding intrinsic symmetries of the system into the model to reduce redundancy. Similar ideas were validated in many machine learning models, e.g., Convolutional Neural Networks (CNNs)\cite{ciresan2011flexible,krizhevsky2012imagenet} suits image recognition better since the respected translation invariance, which also proves efficient in data augmentation of physical systems\cite{Pang:2016vdc, Du:2019civ,Jiang:2021gsw}; In heavy ion collisions, the Point-Net was applied to process point-type particle 
readout from detectors, in which the permutation symmetry was encoded explicitly~\cite{OmanaKuttan:2020brq,Steinheimer:2019iso,OmanaKuttan:2020btb}; In lattice calculations, gauge equivariant flow-based models were designed to handle gauge field configurations sampling~\cite{Kanwar:2020xzo,Boyda:2020hsi,Abbott:2022hkm,Abbott:2022zhs,Abbott:2022zhs}. Besides, the gauge equivariant and invariant neural networks were also proposed for particular quantum systems~\cite{Favoni:2020reg,Namekawa:2021nzu,Luo:2021ccm}. Inspired by the renormalization scheme, the neural network renormalization group was devised also applied to improve MCMC calculations~\cite{li2018neural,Hu:2019nea}.

In this study, we devised a Fourier flow model (dubbed as F-flow in following) for solving imaginary time path integrals with efficient paths generation, whose periodicity is satisfied explicitly since the introduction of a Fourier frequency domain (also known as Matsubara frequency). The model is demonstrated on quantum harmonic and anharmonic oscillator systems. In Section~\ref{sec:path}, we first briefly review the Euclidean Feynman path integral and its periodic condition, which can be tackled with discrete Fourier transform(DFT). In Section~\ref{sec:flow}, we describe the F-flow model in detail.
In Section~\ref{sec:results}, we demonstrate the performance of the proposed F-flow based Feynman path generator on quantum harmonic and anharmonic oscillators. Ground-state wave functions and energy levels up to second excited states are estimated with the F-flow-based path generator. Merits and drawbacks of the F-flow model and its potential applications to other general systems are discussed in the final section.

\section{Euclidean Feynman path integral}\label{sec:path}

Within the path-integral formulation of quantum mechanics, the time evolution of a quantum state, $\psi(x,t)$, can be dictated by the Feynman propagator,
\begin{eqnarray}
\psi(x,t) = \int \mathcal{D}x(t) K(x,t;x_0,t_0) \psi(x_0,t_0),
\end{eqnarray}
where the propagator $K(x,t;x_0,t_0)$(also called kernel) is the sum of all possible paths (or trajectories) connecting the initial point $(x_0, t_0)$ and the end point $(x_f, t_f)$~\cite{feynman2010quantum},
\begin{eqnarray}
K(x,t;x_0,t_0) = A(t-t_0) \sum_{[x(t)]} e^{iS[x(t)]/\hbar}.
\end{eqnarray}
In this functional integral, besides the normalization factor $A(t-t_0)$,  the classical action $S[x(t)]$ appears which includes the kinetic and potential terms. The existence of $i/\hbar$ in the exponent induces quantum fluctuations varying from path to path.  Dramatic fluctuations will induce cancellation for paths when taking the real-time formalism, it's thus convenient to take the Euclidean form of path integrals where the Wick rotation is introduced, $t \to i\tau$, with $\tau$ the imaginary time. The action within a time interval $\mathcal{T}$ in Euclidean space-time derives,
\begin{eqnarray}
\label{action0}
S_E[x(\tau)] &=& \int_0^{\mathcal{T}} d\tau\{T[x(\tau)] + V[x(\tau)] \}
\end{eqnarray}
including the kinetic term $T(x) \equiv \frac{m}{2}(\frac{d x}{d\tau})^2$ with mass $m$, and the potential term $V(x)$. Accordingly, the Euclidean Feynman propagator is,
\begin{eqnarray}
K_E(x,T;x_0,\tau_0) = A_E \sum_{[x(t)]} e^{-S_E[x(\tau)]/\hbar},
\end{eqnarray}
which shows a resemblance to Boltzmann distribution in statistical mechanics and permits Monte Carlo techniques for the evaluation of this integral. From the statistical probabilistic point of view, one can compute any physical observable $\hat{O}$ given all possible paths $\{x(t)\}$ in probability explanation,
\begin{eqnarray}
    \langle \hat{O}\rangle = \int\hat{O}(x)p(x)dx = \sum_{x\sim p(x)} \hat{O}(x),
\end{eqnarray}
where the probability $p(x)$ for each path $x(\tau)$ reads,
\begin{eqnarray}
p[x(\tau)] = \mathcal Z^{-1} e^{-S_E[x(\tau)]/\hbar}\label{eq:pro},
\end{eqnarray}
with the partition function $\mathcal Z = \sum_{x(\tau)} e^{-S_E[x(\tau)]/\hbar}$ as a normalization factor. To connect it to quantum statistical mechanics, one necessitates the periodicity to the path~\cite{Kapusta:2006pm,feynman2010quantum}, thus $x(\tau = 0) = x(\tau = \mathcal{T})$, and views the time interval to be the inverse temperature $\mathcal{T} = \beta$. Taking natural unit $\hbar=1$ and a discrete-time lattice with size $N$, the action accordingly derives as,
\begin{eqnarray}
S_E(\{x_n\}) =\frac{\beta}{N}\sum_{n=0}^{N-1} \bigg[ \frac{m(x_{n+1}-x_n)^2}{2a^2}+V(x_n)\bigg],
\label{action}
\end{eqnarray}
with periodic boundary $x_0 = x_N$ and a discrete time-step $a = \beta/N$, which is adopted as a physical unit($1/a$) for all following calculations. At low temperature, the sites of $x_n$ are strongly correlated, corresponding to $aN \rightarrow \infty$. In all following numerical calculations, we use $a = 0.1$, $N = 100$ to approach this condition.

When enforcing a boundary condition to path integral on the discretized paths $\{x_n\}$, invariance under translation ($\{n\rightarrow n+1\}$), inversion ($\{n\rightarrow -n\}$) and periodicity ($\{n\to n+N\}$) should preserve, where $n$ labels the index of the site. This motivated us to introduce the discrete Fourier transformation(DFT) for the path chain $\{x_n\}$, thus converting the coordinates into Matsubara modes
\begin{eqnarray}
X_k = \sum_{n=0}^{N-1}e^{-i\frac{2\pi}{N}kn}x_n.
\label{eq:dft}
\end{eqnarray}
In this Matsubara space the action derives (see App.~\ref{appa})
\begin{eqnarray}
S(x)\approx \frac{\beta}{N^2} \sum_{k=0}^{N-1} [\frac{m(1-\cos\frac{2\pi k}{N})}{a^2}|X_k|^2 + V(X_k)],\label{eq:kaction}
\end{eqnarray}
which does not couple different Matsubara frequencies.
\begin{figure}[!htb]
	\includegraphics[width=0.35\textwidth]{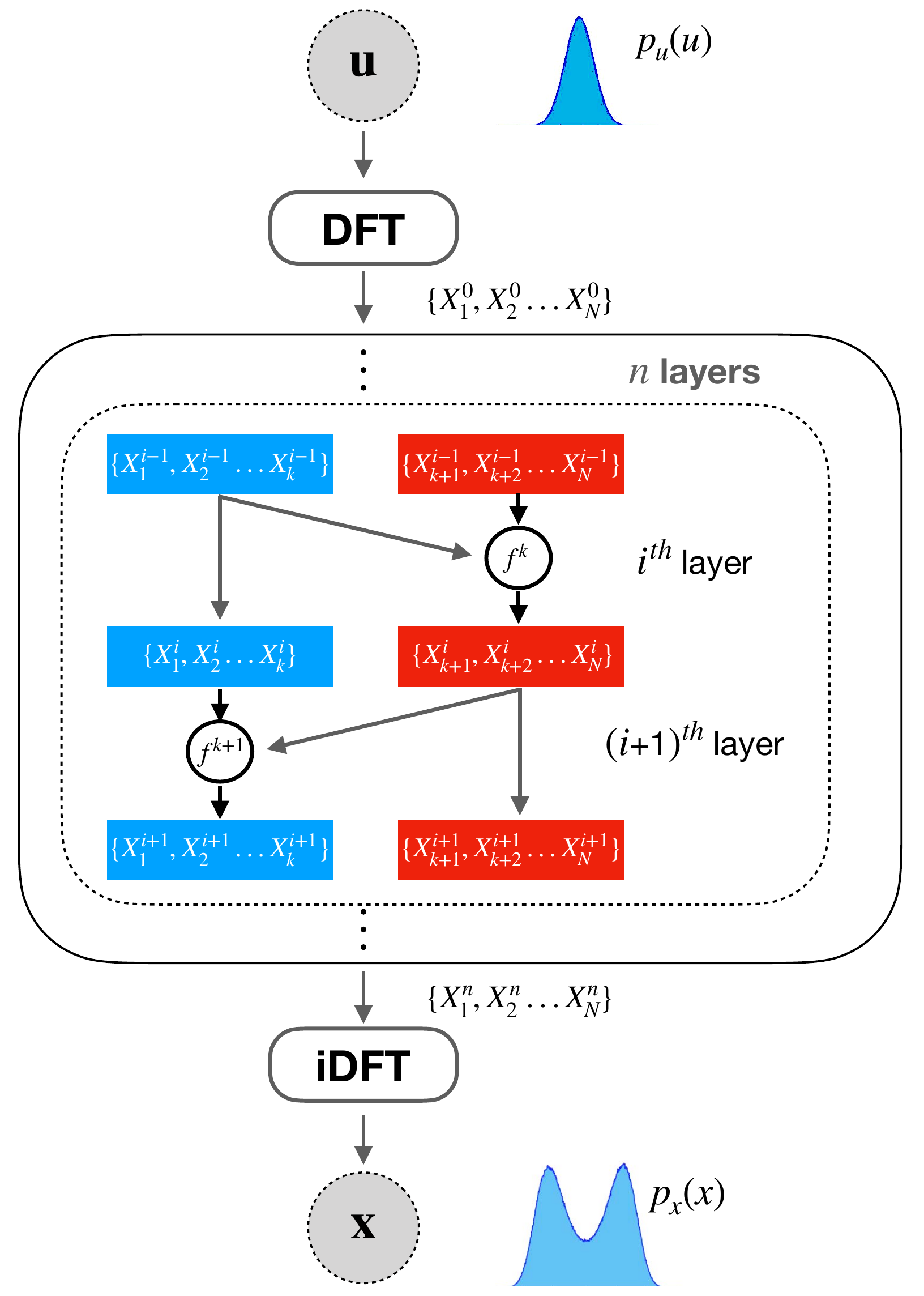} 
	\caption{Structure of the F-flow model with insertion of densely connected layers. The input of the model contains samples generated from a naive prior distribution ${\bf u}\sim p_{\bf u}({\bf u})$, which is a multivariate normal distribution in our case.}
	\label{fig:fflow}
\end{figure}

\section{Fourier-Flow-based path generator}\label{sec:flow}
Flow-based model is proven to be useful for density estimation and interference tasks~\cite{Gao:2020zvv,Gao:2020vdv,Kanwar:2021wzm,Boyda:2022nmh}. In particular, normalizing flows provide a novel way of constructing a flexible probability distribution over continuous random variables~\cite{papamakarios2021normalizing}. The main idea of normalizing flow is to express a complex distribution ${\bf x}\sim p_{\bf x}({\bf x})$ from a naive distribution ${\bf u}\sim p_{\bf u}({\bf u})$ via a bijective transformation $T:{\bf u}\rightarrow {\bf x}$, which could be represented by neural networks with trainable parameters noted as $\{\theta\}$. The probability distribution function changes as follows~\cite{Albergo:2021vyo},
\begin{eqnarray}
p_{\bf x}({\bf x})= p_{\bf u}({\bf u}) |\det J_T({\bf u})|^{-1},
\label{eq:trans}
\end{eqnarray}
where ${\bf u} = T^{-1}({\bf x})$ and the Jacobian matrix of the transformation is $J_T$. The key recipe to construct a feasible flow model is to guarantee the transformation $T$ to be invertible, differentiable and composable. It usually requires a series of transformations to link the naive distribution and the target distribution, denoted as $\{T^{i}\}$.

To construct the F-flow model, we start from the Real NVP (real-valued non-volume preserving) model~\cite{dinh2016density}, where specific affine transformations represented by neural networks rendering accessible Jacobian determinant are adopted. The $i^{\text{th}}$ affine coupling layer reads
\begin{equation}
\left\{
\begin{aligned}
    X^{i}_{1:k} &= X^{i-1}_{1:k} \\
    X^{i}_{k+1:N} &= X^{i-1}_{k+1:N} \odot e^{s_{\theta}^i(X^{i-1}_{1:k})} + t_{\theta}^i(X^{i-1}_{1:k}),
\end{aligned}
\right.
\end{equation}
with the subscript labels $N$ nodes at $i^{\text{th}}$ layer and ``$\odot$'' represents the element-wise product. The neural networks are used to construct mappings $s_{\theta}: \mathbb{R}^{k}\rightarrow \mathbb{R}^{N-k}$ and $t_{\theta}: \mathbb{R}^{k}\rightarrow \mathbb{R}^{N-k}$ for scaling and translation transformations with $\{\theta\}$ denoting the network parameters. 

Based on the above affine transformation for each coupling layer, one can calculate directly the determinant of Jacobian by tracing a lower triangular matrix, $(\det J_{T}^i) = \Pi_j^{N-k}e^{s_{\theta}^i(X_{1:k})_j}$.

Fig.\ref{fig:fflow} depicts the devised F-flow model in this work, where the DFT (Eq.~\eqref{eq:dft}) is introduced before the first affine coupling layer, and the inverse DFT (iDFT) is performed after the last affine coupling layer to convert paths from frequency space to coordinate space, $x_n =\frac{1}{N} \sum_{k=0}^{N-1}e^{i\frac{2\pi}{N}kn}X_k$. This ensures that all transformations in the flow model are in frequency space, which on one hand preserves the periodicity automatically, and on the other hand explicitly includes all quantum fluctuations since all the relevant modes are transformed and generated inside the flow.

To approach the desired probability distribution $p(x)$ shown in Eq.\eqref{eq:pro}, one should define a proper loss function to tune the parameters of the neural networks in the flow model. Kullback-Leibler(KL) divergence provides a natural measure for the dis-similarity between two distributions~\cite{mezard2009information}, and thus it's taken to define the loss function for the F-flow model training,
\begin{eqnarray}
\mathcal L(\theta) &=& D_{KL} [p_{\bf x}({\bf x};\theta)||p({\bf x})]\nonumber\\
 &=& \mathbb E_{{\bf u}\sim p_{\bf u}({\bf u})}[\log p_u({\bf u})-\log|\det J_T({\bf u})|]\nonumber\\
 &&- E_{{\bf x}\sim p_{\bf x}({\bf x})}\log p({\bf x})
\end{eqnarray}
where $p(x) \equiv p({\bf x)}$ is the target distribution (here Eq.~\ref{eq:pro}), and $q_\theta(x)\equiv p_{\bf x}({\bf x};\theta) = p_{\bf u}({\bf u})|\det J_T({\bf u})|^{-1}$ the F-flow parameterized distribution via $x = T({\bf u})$. Introducing the path integrals, then the loss function derives as, 
\begin{eqnarray}
\mathcal L(\theta) &=& \mathbb E_{x\sim q_\theta(x)} [ S_E(x) + \ln q_\theta(x)] + \ln \mathcal{Z}.
\end{eqnarray}
The partition function contributes as a constant term which does not bring gradients to trainable parameters. We thus omit the last term in the training process.

\section{Generating Feynman Paths and Estimating Observables}~\label{sec:results}
We first demonstrate the proposed F-flow model on the quantum harmonic oscillator, with analytical solutions existing. After that, we move to the harmonic oscillator, where no analytical methods yet are applicable.
\subsection{Harmonic oscillator}
The typical harmonic potential reads,
\begin{eqnarray}
V(x) = \frac{1}{2}\mu x ^2,
\end{eqnarray}
with $\mu$ the coupling constant. The energy levels can be analytically solved as $E_n = \frac{n+1}{2}\mu$, in which the ground state is treated as the zero-point energy $E_0 = \frac{1}{2}\mu$. Here we set $\mu=1$ and $m=1$ for demonstration. In the training process, we set $N_{sample}=8192$ for the model to evaluate the loss gradient, and set the iteration of each epoch to be $N_{iter} = 1024$ with total $N_{epoch} = 10$. Note that the generation of path is uncorrelated by definition and efficient enough after the training: 0.0015 seconds for generating 1000 samples on single game GPU (RTX3090).

\begin{figure}[!htbp]
	\includegraphics[width=0.4\textwidth]{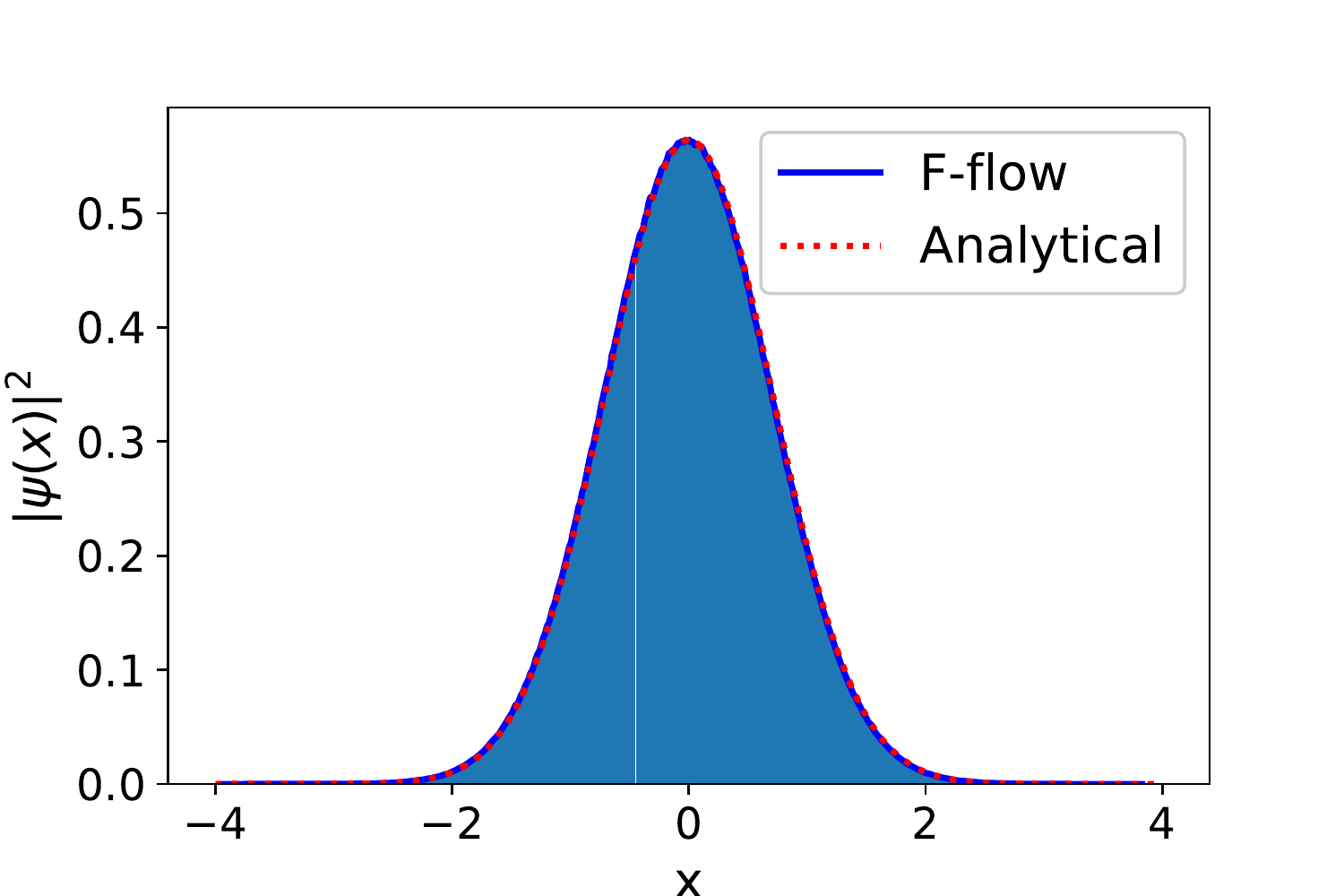} 
	\caption{The ground state wave function square for harmonic oscillator, with red dashed curve the analytical results, and the blue curve evaluated from F-flow model.}
	\label{harmonicswf}
\end{figure}

From the Virial theorem, $2\langle T \rangle = n\langle V\rangle$ holds for each potential term with n the power of $x$, and we get the ground state energy of the harmonic oscillator to be $E_0=\mu^2\langle x^2 \rangle$.
Since the simulation is under the low temperature limit, which mean the sites are strongly correlated, we could push our calculation for the two-points and four-points correlation to obtain the energy of the first and second excited state,
\begin{eqnarray}
E_1-E_0 &=& -\lim_{\tau\rightarrow \infty}\frac{d \log G_2(\tau)}{d\tau},\nonumber\\
E_2-E_0 &=& -\lim_{\tau\rightarrow \infty}\frac{d \log G_4(\tau)}{d\tau},
\label{eq:excited}
\end{eqnarray}
with $G_2 = \lim\limits_{T\rightarrow \infty}(\langle x(\tau)x(0)\rangle - \langle x(\tau)\rangle \langle x(0)\rangle)$ and $G_4 = \lim\limits_{T\rightarrow \infty}(\langle x(\tau)^2x(0)^2\rangle - \langle x(\tau)^2\rangle \langle x(0)^2\rangle)$ the two-points and four-points correlation functions, respectively.
\begin{table}[!htbp]
    \centering
    \begin{tabular}{cccc}
    \hline \hline
    Energy& Analytical &F-flow(200k) & F-flow(400k)\\
    \hline
    $E_0$& 0.5 & 0.4997& 0.4997\\
    \hline
    $E_1$& 1.5 & 1.5171 & 1.4999\\
    \hline
    $E_2$& 2.5 & 2.5324 & 2.5019\\
    \hline \hline
    \end{tabular}
    \caption{Quantum harmonic oscillator energy levels from the analytical and F-flow model estimation (take $N=100$ for the discrete time lattice). The difference between the two F-flow model results lies in the number of samples taken.}
    \label{tab:harm}  
\end{table}
In Table.~\ref{tab:harm}, we summarize results of F-flow model with a comparison to analytical results for low-lying energy levels up to the second excited state.
The estimation with F-flow model agrees very well with the analytical results. For the F-flow model evaluation, we sampled 200k and 400k paths to estimate the energy level, which clearly indicates that increasing statistics from F-flow model converge better to analytical results. The ground state wave function square $|\psi(x)|^2$ can be obtained by evaluating the probability density for finding a particle in the interval ($x\sim x+dx$) from the sampled paths. Fig.~\ref{harmonicswf} shows the wave function square evaluated from F-flow model, which matches perfectly with the analytical result. 
\subsection{Anharmonic oscillator}
\begin{figure}[!htbp]
	\includegraphics[width=0.52\textwidth]{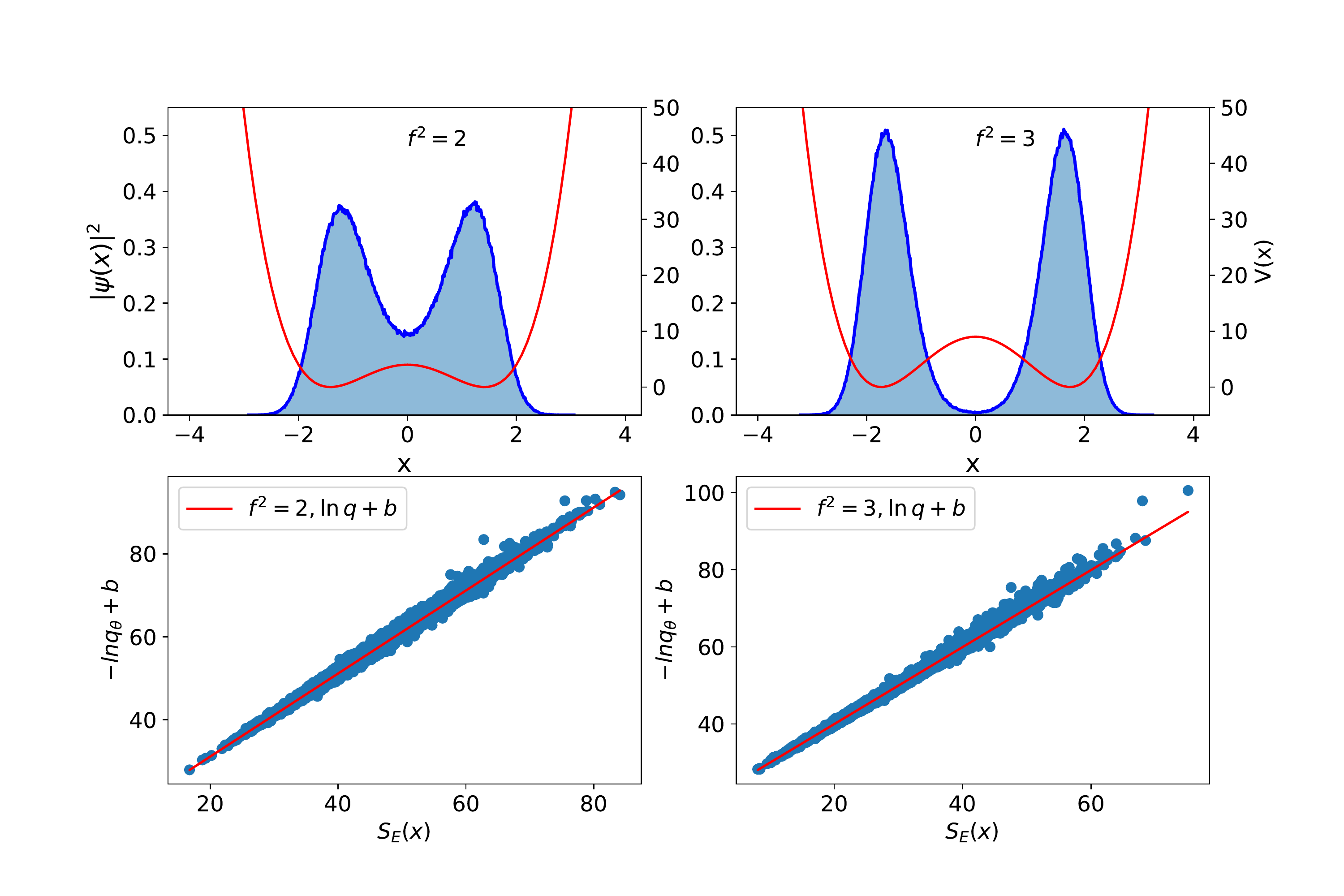} 
	\caption{(Upper)The ground state wave function square obtained from F-flow model for anharmonic oscillator with $f^2 = 2$ (left) and $f^2 = 3$ (right). And the red lines are the corresponding potantial. (Lower) The correlation between the effective action $S_{eff}$ from F-flow model and the true action.}
	\label{wavefunction}
\end{figure}

After the above verification, we implement the F-flow model to anharmonic oscillator. We consider the following one-dimensional double-well potential~\cite{Blankenbecler:1979pa,Stevenson:1981vj,Jansen:2013jpa},
\begin{eqnarray}
V(x) = \lambda (x^2-f^2)^2 ,
\end{eqnarray}
which under DFT turns to
\begin{eqnarray}
V(q) = \lambda \sum_{k_1,k_2}\frac{1}{N^2}X_{-k_1}X_{-k_2}X_{k_1-q}X_{k_2+q}\nonumber\\
-2\lambda f^2 \sum_{k_1} X_{k_1}X_{-k_1+q}+\lambda f^4 .
\end{eqnarray}
After DFT, as shown by Eq.~\eqref{eq:kaction}, correlation between $x_n$ in the kinetic term disappears. Instead, one has a collection of modes $X_k$ correlated via potential $V(q)$. It's worth noting that when removing DFT and iDFT for the model here (i.e., degrade to plain normalizing flow), it fails for the path integral of anharmonic oscillator, which was also observed recently in the literature~\cite{Che:2022gzu,Hackett:2021idh}.

For numerical simulation, we choose $a = 0.1$,  $N = 100$ (for the discrete time lattice), $m = 0.5$, $\lambda = 1$. The same training hyperparameters as in harmonic oscillator is taken here for the anharmonic oscillator simulation, and seven situations with $ f^2 = \{5,4,3,2,1,0,-1\}$ are simulated with which we estimated the ground state wave function and energy levels up to second excited state. The ground state energy of double-well potential can be derived from the Virial theorem as,
\begin{eqnarray}
E_0 = 3\lambda \langle x^4\rangle - 4\lambda f^2 \langle x^2\rangle +\lambda f^4.
\end{eqnarray}
Again, Eq.~\eqref{eq:excited} is used for estimating energy levels for excited states. 
From Fig.~\ref{wavefunction}(upper panel) , it's obvious that with increasing $f^2$ the potential deviates farther away from harmonics, and shows a higher potential barrier between two wells, where the solution of the system is nontrivial due to the involved tunneling through the barrier.

Fig.~\ref{wavefunction} (upper panel) also shows the ground state wave function square estimated from F-flow, to which a double peak structure due to quantum tunneling appeared. We found that without the introduction of frequency domain the flow evaluation easily collapse to single peak wave function. The explicit operation and inclusion over all Matsubara mode from Fourier transformation in our model safely brings in the needed ``tunnelling'' events (see App.~\ref{appb} for typical paths sampled from F-flow) for the double peak wave function, which achieves efficient multi-modal distribution sampling. It's also seen that with increasing $f^2$ the overlap between the two peaks in the wave function would decrease, meanwhile each peak in the wave function shrinks. Fig.~\ref{wavefunction} (lower panel) show the correlation between the effective action (i.e., $- \log q_{\theta}(x)$ by the F-flow) and the true action $S_E(x)$. We see that the F-flow captured effective action closely resembles the true action after accounting for an overall constant shift.

\begin{figure}[!htb]
	\includegraphics[width=0.5\textwidth]{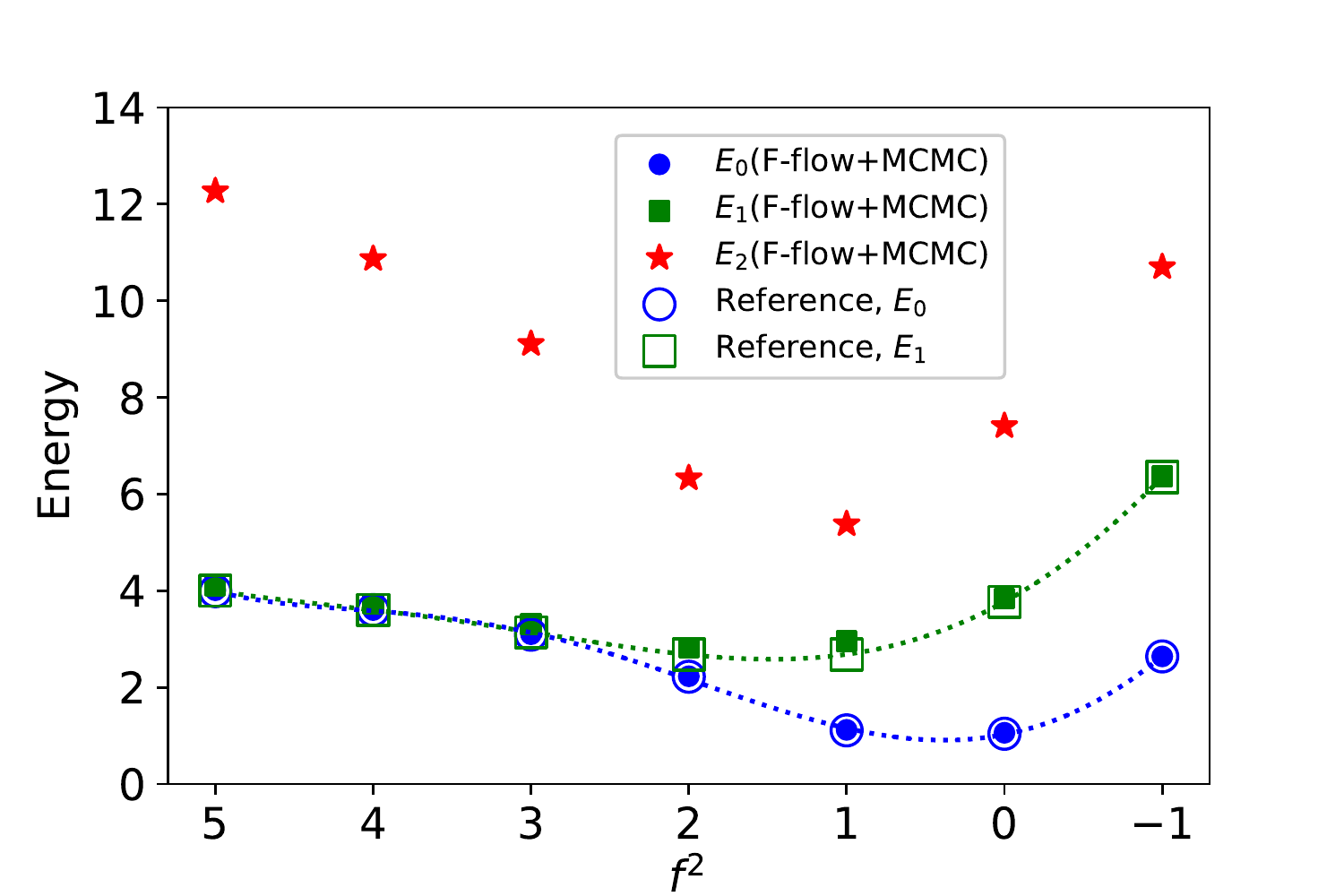} 
	\caption{The first three energy levels of the anharmonic oscillator as function of $f^2$. The solid markers are from F-flow model calculation, and the hollow ones with also dashed curve are from the continuum theory~\cite{Blankenbecler:1979pa}.}
	\label{energylevel}
\end{figure}

\begin{table}[!h]
\caption{Low-lying energy levels from F-flow model (upper), F-flow augmented MCMC (middle) and MCMC (lower).}
\label{energywithMCMC}  
\centering
\begin{tabular}{cccccccccc}
\\
\hline \hline
\multicolumn{3}{c}{$f^2$} -1.0 & 0.0 & 1.0 & 2.0 & 3.0 & 4.0 & 5.0 \\
 \hline
\multirow{3}{*}{\textbf{$E_0$}}
& F-flow & 2.66 & 1.06 & 1.15 &2.28        & 3.14   & 3.60    & 4.03 \\
& F-flow+MCMC & 2.64 & 1.04 & 1.11 & 2.23   & 3.10   & 3.60    & 4.01 \\
& MCMC & 2.64 & 1.04 & 1.11 & 2.23        & 3.10   & 3.60 & 4.00 \\
 \hline
\multirow{3}{*}{\textbf{$E_1$}}
& F-flow & 6.37 & 3.84 & 2.96 & 2.87       & 3.38   & 3.74    & 4.13 \\
& F-flow+MCMC & 6.35 & 3.76 & 2.74 & 2.82  & 3.31   & 3.73    & 4.10 \\
& MCMC & 6.35 & 3.77 & 2.71 & 2.82       & 3.31   & 3.73    & 4.09 \\
 \hline
\multirow{3}{*}{\textbf{$E_2$}}
& F-flow & 10.7 & 7.41 & 5.38 & 6.36       & 9.12  & 10.91 & 12.30  \\
& F-flow+MCMC & 10.69 & 7.44 & 5.87 & 6.33  & 9.11  & 10.87 & 12.27  \\
& MCMC & 10.68 & 7.41 & 5.85 & 6.35       & 9.08  & 10.85 & 12.22  \\
\hline \hline
\end{tabular}
\end{table}
Fig.~\ref{energylevel} and Table.~\ref{energywithMCMC} display the evaluated low-lying energy levels, including the ground state, first and second excited state energy for different values of the parameter $f^2$. Results solely from F-flow model already agree well with results from MCMC evaluation. In the table the MCMC results are obtained using the method outlined in \cite{Creutz:1980gp}, $12\times10^6$ configurations were sampled in total of which $6\times10^5$ used to calculate energy of the ground and exited states. F-flow also gives exactly the same tendency across different states and different values of $f^2$ compared with MCMC estimations. By taking the trained F-flow model as a proposal in a Markov Chain process we get closer results as compared to pure MCMC, while the autocorrelation time is significantly reduced (see App.\ref{autotime}). Note that on average the acceptance rate using F-flow for proposal in MCMC is always above $50\%$ since the uncorrelated sampling from the trained F-flow. These all demonstrate that the proposed F-flow-based generative model is valid in constructing a more efficient Feynman path generator for the quantum system.

\section{Summary}
Feynman path integral provides an intuitive and the most fundamental way to represent quantum evolution and dynamics. Since the track ergodicity, it's crucial to be able to generate Feynman's path efficiently according to its probability distribution derived from path integral. From MCMC\cite{Cranmer:2019kyz} to VAE\cite{Che:2022gzu}, there have been many efforts taken trying to make this process more accurate and efficient. In this work, we propose a Fourier-flow model involving DFT and generative Real NVP method to render a much less time-costing and automatic symmetry preserving Feynman's path generator, with which all the quantum information including the evolution propagator, ground state wave function and low-lying energy levels can be evaluated efficaciously. 

The demonstration of the proposed F-flow model on quantum harmonic and anharmonic oscillators shows its applicability and success for investigating quantum systems by efficient Feynman's path generation. Compared to conventional MCMC, the proposed F-flow model gives more efficient path generation due to the uncorrelation nature and inherent parallel manner for the sampling inside the model. Taking the F-flow model as a proposal on a Markov Chain, we can get a general and effectively augmented MCMC approach with significantly boosted efficiency, where also the exact path generator is guaranteed mathematically. 

It's worth mentioning that the introduction of the Fourier-flow model in this work is motivated by observing the failure of the pure normalizing flow model in achieving the typical symmetry (boundary condition) for the quantum system, which is an eternal theme in physics. Many recent researches have made important contributions to such a problem like generating gauge field configurations\cite{Kanwar:2020xzo,Albergo:2019eim}, and, our work for the first time from Fourier (Matsubara) space point of view to pave the way in tackling this problem, where the affine transformation inside flow model learn to approach a Matsubara representation instead. We will explore further the application of the method for studying other quantum statistics and dynamics problems, to QFT as well, by a combination of state-of-the-art deep learning strategies with physics priors.

\section*{Acknowledgement}
We thank Drs. Shuzhe Shi and Partha Bagchi for helpful discussions. The work is supported by (i) the BMBF under the ErUM-Data project (K. Z.), (ii) the AI grant of SAMSON AG, Frankfurt (K. Z. and L. W.), (iii) Xidian-FIAS International Joint Research Center (L. W), K. Z. also thanks the donation of NVIDIA GPUs from NVIDIA Corporation.
\newpage
\bibliography{ffm}
\bibliographystyle{apsrev4-1}

\appendix
\section{Euclidean Feynman path integral in Fourier space}
\label{appa}
All the mentioned invariances in the end of Sec.~\ref{sec:path} can be preserved explicitly when we transform paths from the coordinate space to the frequency space, and correspondingly the Euclidean action derives as
\begin{eqnarray}
S_E[x(\tau)]  = \frac{\beta}{N} \sum_{n=0}^{N-1}[\frac{m}{2a^2}(x_{n+1} - x_n)(x_{n+1} - x_n)^*  + V(x_n)]\nonumber\\\label{eq:action}
\end{eqnarray}
with its first term becomes
\begin{eqnarray}
&&\sum_{n=0}^{N-1}(x_{n+1} - x_n)(x_{n+1} - x_n)^*\nonumber\\
&=& \sum_{n=0}^{N-1}(x_{n+1}^2+x_n^2) - \sum_{n=0}^{N-1}(x_{n+1}^*x_n + x_n^*x_{n+1})\nonumber\\
&=& \frac{2}{N}\sum_{k=0}^{N-1}|X_k|^2 - \sum_{n=0}^{N-1}(x_{n+1}^*x_n + x_n^*x_{n+1})\nonumber\\
&=& \frac{2}{N}\sum_{k=0}^{N-1}|X_k|^2 - \frac{1}{N}\sum_{k'=0}^{N-1}(e^{i\frac{2\pi}{N}k'} + e^{-i\frac{2\pi}{N}k'})|X_{k'}|^2\nonumber\\
&=& \frac{2}{N}\sum_{k=0}^{N-1}|X_k|^2(1-\cos\frac{2\pi k}{N})
\end{eqnarray}
where the unitary of the Fourier transformation is used to reduce the formula among the first three equations. Take the last line back to Eq.~\eqref{eq:action}, one can get,
\begin{eqnarray}
S_E[x(\tau)] = \frac{\beta}{N} [\frac{1}{N}\sum_{k=0}^{N-1}\frac{m(1-\cos\frac{2\pi k}{N})}{a^2}|X_k|^2 + \sum_{n=0}^{N-1} V(x_n)],\nonumber\\
\end{eqnarray}
and then after the DFT for the potential term, the action can be derived as, 
\begin{eqnarray}
S(x)\approx \frac{\beta}{N^2} \sum_{k=0}^{N-1} [\frac{m(1-\cos\frac{2\pi k}{N})}{a^2}|X_k|^2 + V(X_k)]. \nonumber \\
\label{eq:2kaction}
\end{eqnarray}

\section{Autocorrelation time}
\label{autotime}
We calculate the correlation function of variable $\langle X\rangle$ of MCMC process based on the F-flow trained configuration dataset with the final autocorrelation time $\tau = 0.42$. This result shows that the F-flow procedure can remarkably make the traversing process more efficient.
\begin{figure}[!htbp]
	\includegraphics[width=0.4\textwidth]{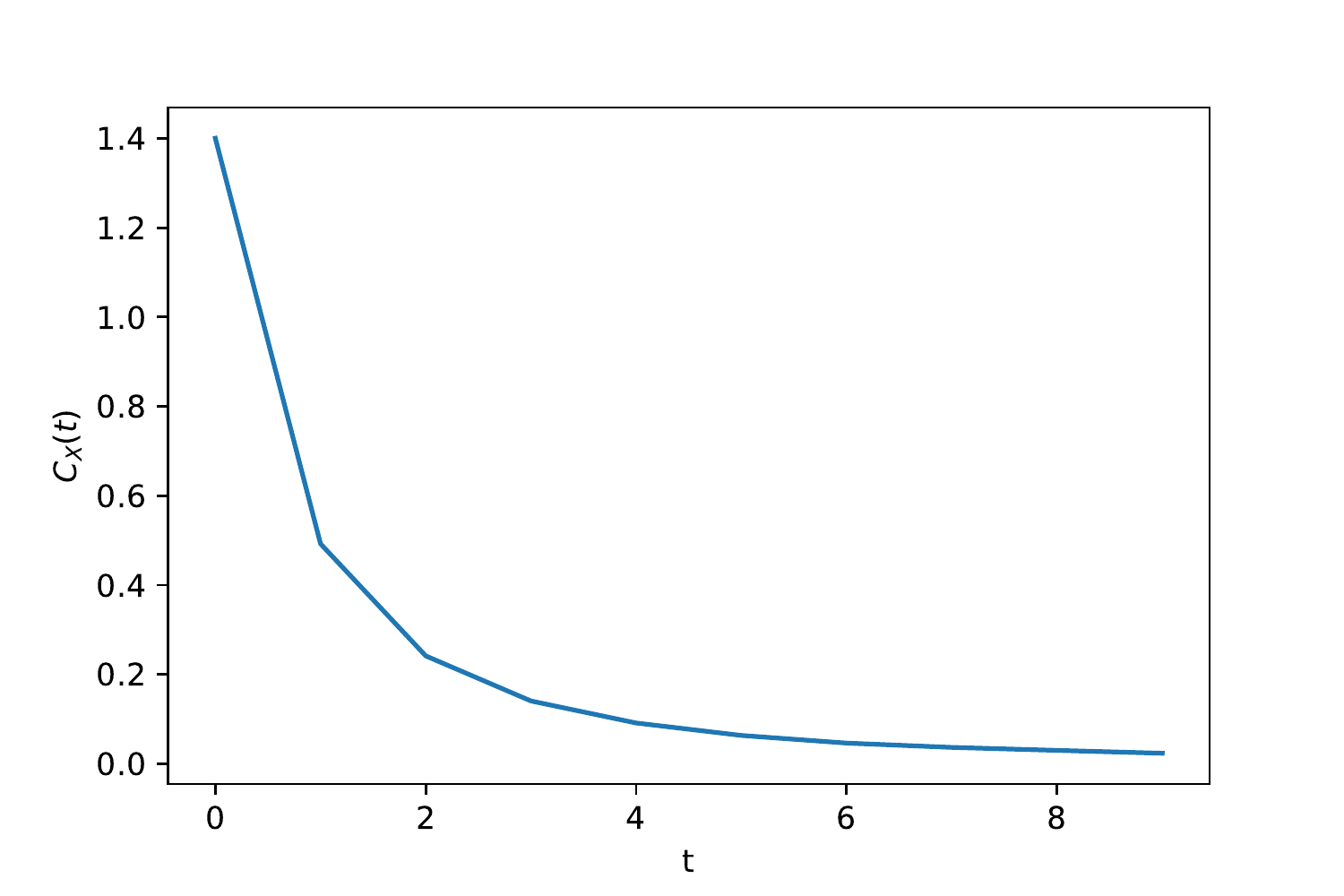} 
	\caption{The correlation function of path configuration versus discrete time of Markov chain.}
	\label{auto}
\end{figure}

\section{Anharmonic oscillator paths}
\label{appb}
\begin{figure*}[!htbp]
    \centering
    \begin{minipage}{0.4\linewidth}
    \centering
    \includegraphics[width=0.9\textwidth]{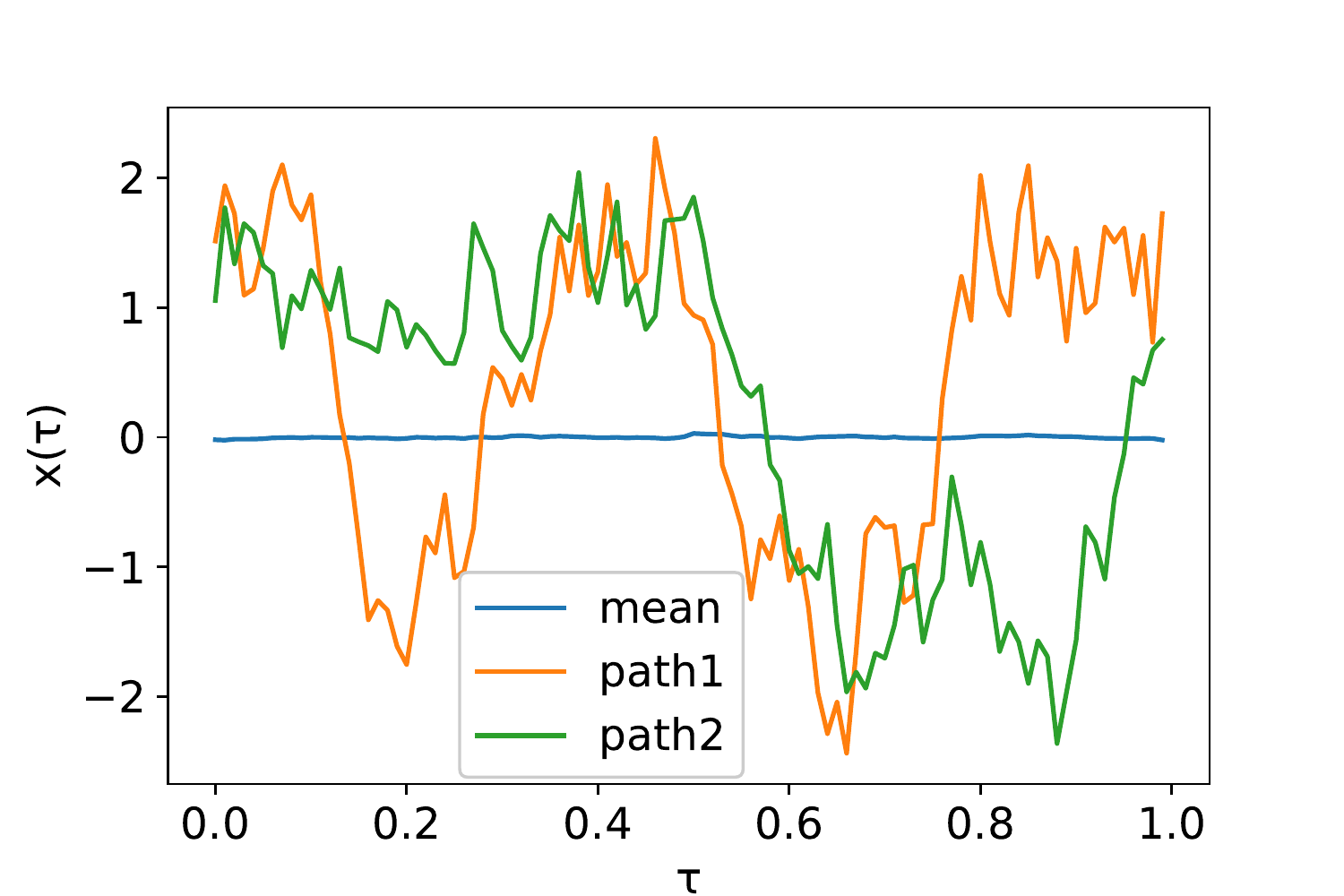} 
    \end{minipage}
    \begin{minipage}{0.4\linewidth}
    \centering
    \includegraphics[width=0.9\textwidth]{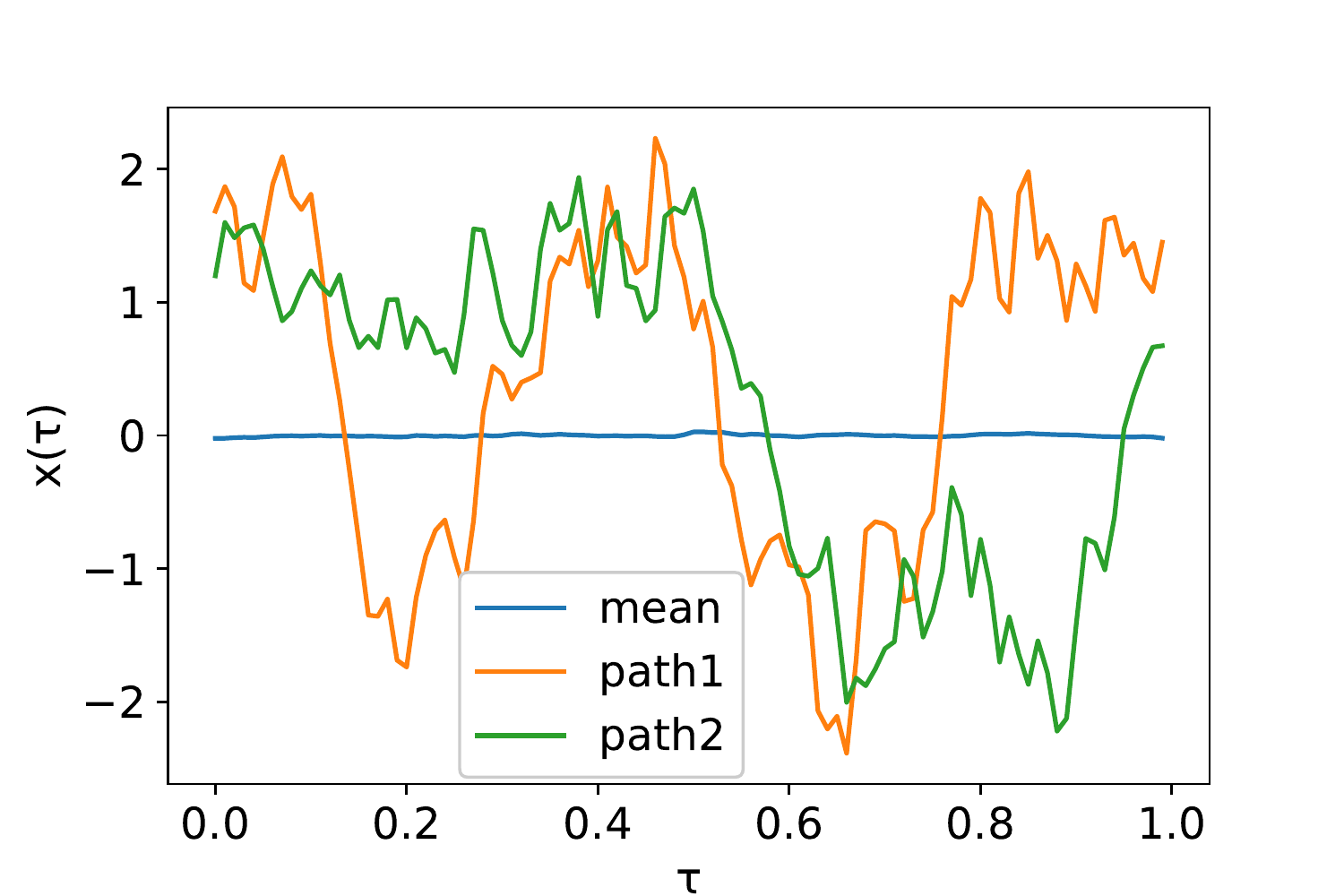} 
    \end{minipage}
    \begin{minipage}{0.4\linewidth}
    \centering
    \includegraphics[width=0.9\textwidth]{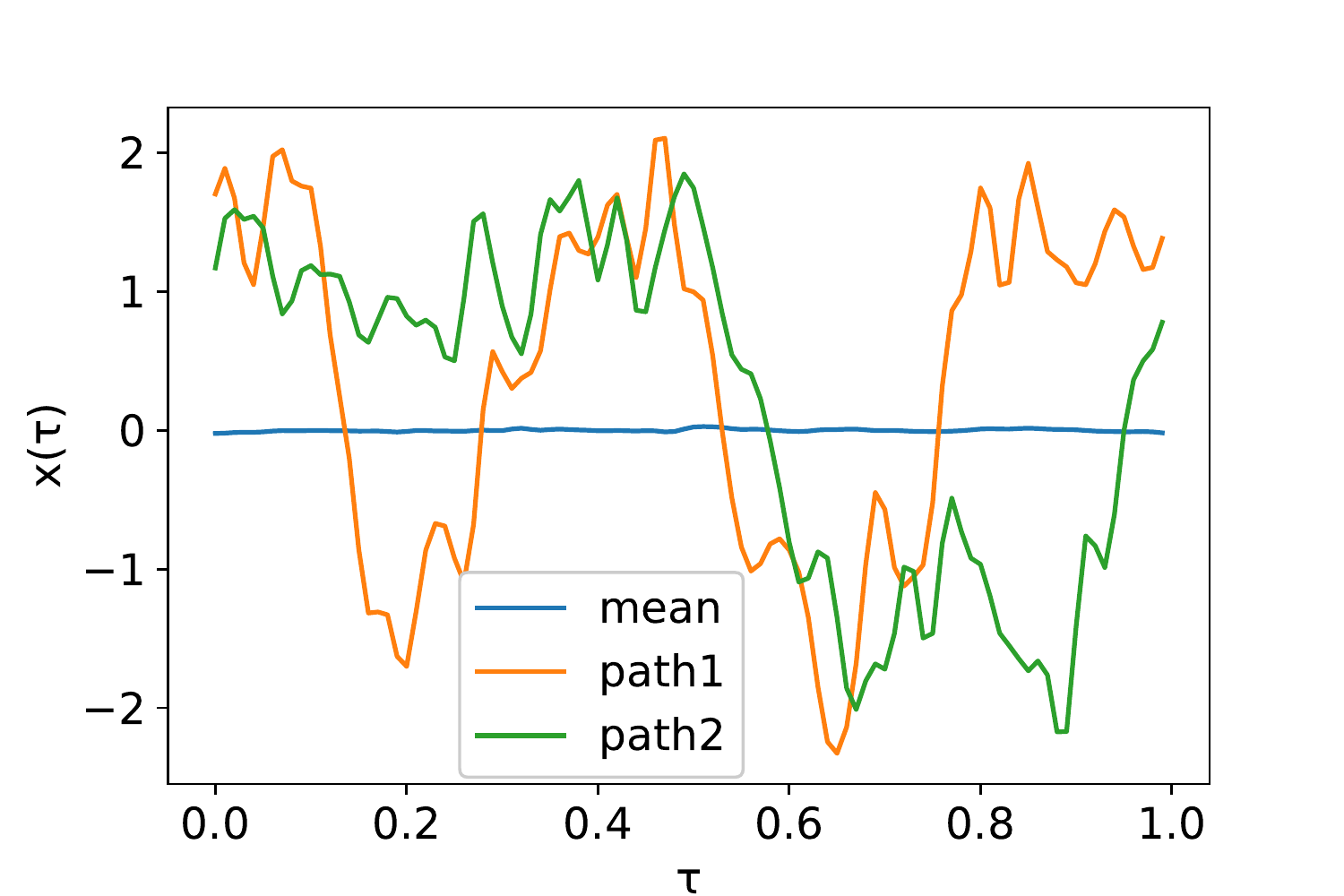} 
    \end{minipage}
    \begin{minipage}{0.4\linewidth}
    \centering
    \includegraphics[width=0.9\textwidth]{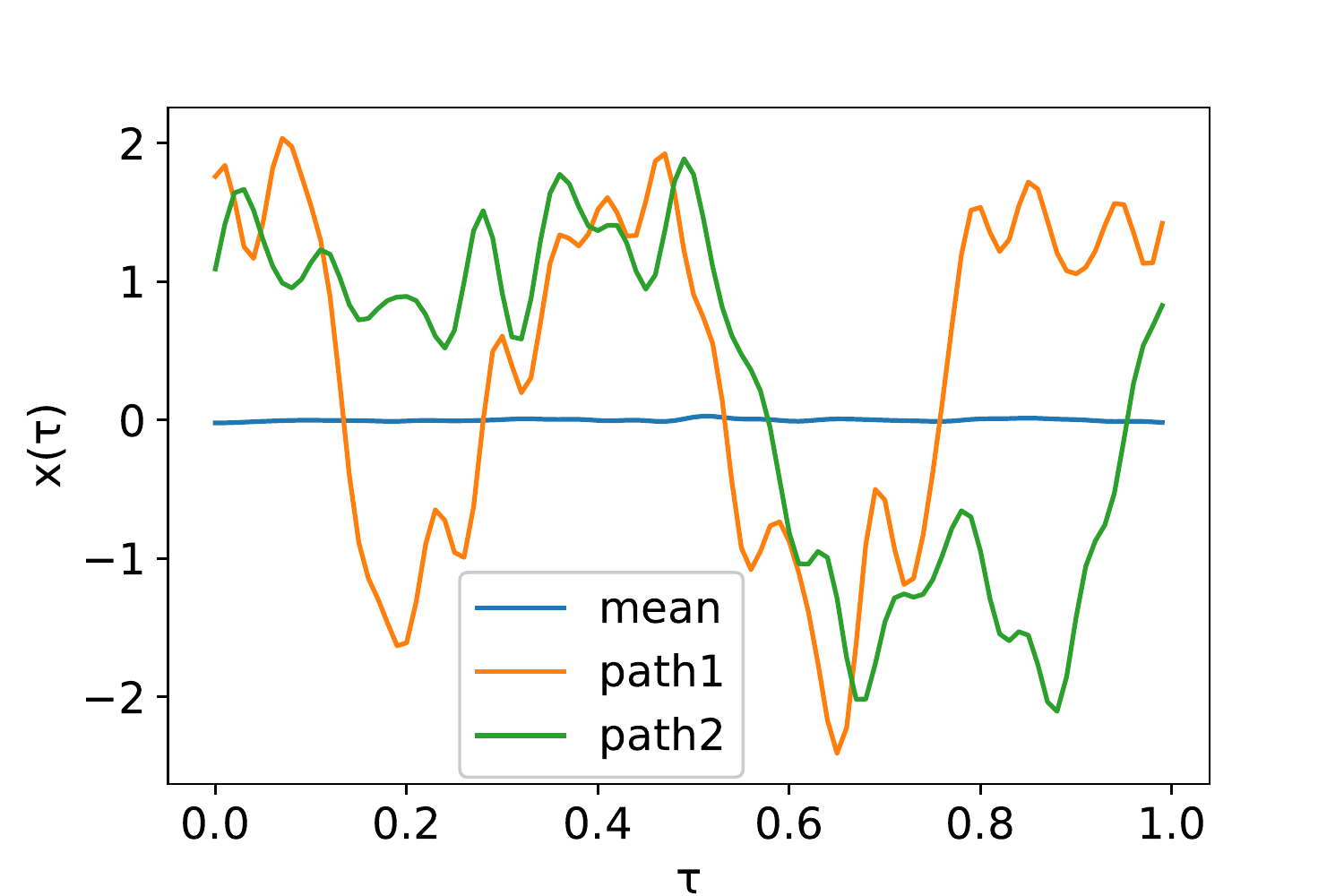} 
    \end{minipage}
	\caption{Typical Feynman paths for quantum anharmonic oscillator with $f^2=2$ sampled from the trained F-flow model. Upper-left is the original paths sampled with full frequency modes used. From upper-left to bottom-right is gradually masking more high frequency modes (0 \%, 20\%, 30\%, 40\%) after DFT then reverse via iDFT back to path.}
	\label{paths}
\end{figure*}
Typical Feynman paths from our trained F-flow model on the anharmonic oscillator are shown in Fig.~\ref{paths} (upper left). For exploration, it's interesting to see what if one mask (i.e., set to zeros) high-frequency modes would influence the path, since if this would not induce much difference one could largely reduce the flow transformation's parameters in frequency space for the path generation.
In our proposed F-flow model, because we insert the discrete Fourier transformation before and after the affine transformation (represented by networks), we wonder whether the low-frequency modes can dominate the generated paths. When we mask the high-frequency modes in frequency space and then reverse via iDFT back to the coordinate space, as shown in Fig.~\ref{paths}, we find that the overall shape of paths does not change but just gets smoothed (i.e., the fluctuation within small areas disappears). Without the high-frequency mode, we can reduce the training parameters to save time and cost. Note that one should further investigate in detail the influence on physical observables from the mask of high-frequency modes, which is left for future exploration.
In principle, we may not regard this process as an appropriate way to handle the path integral problem because those high frequency modes also contribute to the energy of the system. But, this may be an inspiration to generative tasks in computer vision problem, like to speed up the generation or for super-resolution development.

\end{document}